\documentclass[12pt,english]{article}
\usepackage[T1]{fontenc}
\usepackage[latin9]{inputenc}
\usepackage{amsmath}
\usepackage{amssymb}
\usepackage{graphicx}

\makeatletter
\usepackage{jheppub}
\def\@fpheader{\relax}
\allowdisplaybreaks[1]

\makeatother

\usepackage{babel}
\begin{document}
\subheader{}


\title{Modified entropies as the origin of generalized uncertainty principles}
\author[a]{Nana Cabo Bizet,}
\author[a]{Octavio Obreg\'on,}
\author[a]{Wilfredo Yupanqui.}
\affiliation[a]{Departamento de F\'isica, Divisi\'on de Ciencias e Ingenier\'ias, Universidad de Guanajuato
Loma del Bosque 103, Le\'on 37150, Guanajuato, M\'exico.}
\emailAdd{nana@fisica.ugto.mx}
\emailAdd{octavio@fisica.ugto.mx}
\emailAdd{w.yupanquicarpio@ugto.mx}

\abstract{The Heisenberg uncertainty principle is known to be connected to the entropic uncertainty principle. This correspondence is obtained employing a Gaussian probability distribution for wave functions associated to the Shannon entropy.  Independently, due to quantum gravity effects the Heisenberg uncertainty principle has been extended to a Generalized Uncertainty Principle (GUP). In this work, we show that GUP has been derived from considering non-extensive entropies, proposed by one of us. We found that the deformation parameters associated with $S_{+}$ and $S_-$ entropies are negative and positive respectively.  This allows us to explore various possibilities in the search of physical implications. We conclude that non-extensive statistics constitutes a signature of quantum gravity.
}
\maketitle

\section{Introduction}
Physical phenomena on the smaller scales are described by Quantum Mechanics, and our world is inherently non deterministic according to this theory. Heisenberg's uncertainty principle \cite{heisenberg},  implies that it is impossible to have a particle for which both position $q$ and momentum $p$ are sharply defined, in dramatic contrast with classical mechanics. This principle is summarized in the inequality $\sigma(q) \sigma(p) \geq 1/2$ (considering $\hbar=1$), which was proposed by  Kennard \cite{kennard}. The quantities $\sigma(q)$ and $\sigma(p)$ are defined as standard deviations of position and momentum respectively. The standard deviation is used a lot in the statistical analysis of experiments. It is a reasonable measure of the spread when the distribution in question is of a simple "hump" type. In particular, it is a very good characteristic for a Gaussian distribution since it measures the half-width of this distribution. However, when the distribution of the values has more than one hump or when it is not of any simple type, the standard deviation loses some of its usefulness, especially in connection with the notion of uncertainty.

A more appropriate measure of uncertainty can be found by connecting quantum mechanics and statistical mechanics. The connection between quantum mechanics and statistical mechanics is a fruitful arena, in which interesting features arise \cite{baez}. In particular, the Feynman path integral formulation for the partition function in quantum mechanics can be applied as well to explore partition functions in statistical mechanics. It is also noticeable that the uncertainty principle in Quantum Mechanics can be linked to an Entropy Uncertainty Principle. This has been seen in many contexts.

In recent years it has been argued that a more appropriate measure of uncertainty is information entropy \cite{BIALYNICKIBIRULA1984253}. In this work, it is argued that the standard deviation, is not the best measure of uncertainty. But, is possible to overcome the deficiencies of the traditional approach to uncertainty relations while still keeping intact the spirit of the Heisenberg ideas, using a very profound definition of uncertainty, such definition comes from information theory.

This fundamental uncertainty relation is presented in reviews and articles, taking a different perspective (see page 5 of Coles \cite{coles15}). This review surveys entropic uncertainty relations that capture Heisenberg's idea, which states that the results of incompatible measurements are impossible to predict, covering both finite and infinite-dimensional measurements. These ideas are then extended to incorporate quantum correlations between the observed object and its environment, allowing for a variety of recent, more general formulations of the uncertainty principle. In particular, Coles and co-authors showed in a pedagogical calculation how we can deduce the Heisenberg uncertainty relation, from the entropic uncertainty relations, assuming a Gaussian probability distribution \cite{coles15}.

We will now discuss a generalization of Heisenberg's uncertainty relation, which is commonly known as the Generalized Uncertainty Principle (GUP) that was first obtained by Veneziano et al \cite{AMATI198941}. They studied the scattering of strings at very high energies in order to analyze the inconsistencies of quantum gravity at the Planck scale. Others very interesting works are due to Scardigli \cite{SCARDIGLI199939} and to M. Maggiore \cite{MAGGIORE199365}; the last author obtained a GUP expression by proposing a Gedankenexperiment for the measurement of the area of the apparent horizon of a black hole in quantum gravity. This independent approach results in a GUP that agrees in its functional form with the result obtained in string theory. The idea of GUP arises from the existence of a minimum measurable length \cite{Garay}, associated with minimal uncertainty in position measurements. In fact, such a minimal length arises in different contexts, for example in string theory \cite{GROSS1988407,AMATI198941}, loop quantum gravity \cite{ROVELLI1995593,Bosso_2021}, and thought experiments of black hole physics \cite{MAGGIORE199365,SCARDIGLI199939}. These common characteristics of several theories of quantum gravity led to a phenomenological model consisting of a modification of the uncertainty principle. In one spatial dimension the simplest generalized uncertainty relation which implies a nonzero minimal uncertainty $\Delta x_0$ in the position has the form \cite{PhysRevD.52.1108}
\begin{eqnarray}
\Delta x\Delta p\geq\frac{1}{2}\left[1+\alpha\left(\Delta p\right)^2\right],\label{Eqq.GUP.1}
\end{eqnarray}
where $\alpha=\alpha_0/M_{Pl}$ is called the GUP parameter or deformation parameter and $M_{Pl}$ is the Planck mass scale. The commutation relation for this generalized uncertainty relation is given by
\begin{eqnarray}
\left[\hat{x},\hat{p}\right]=i\left(1+\alpha\  p^2\right).\label{Eqq.GCR.1}
\end{eqnarray}
A version of this model considers a modification of the Heisenberg algebra \cite{PhysRevD.52.1108,Bosso.PhysRevD.97.126010,Venegas.Elias} to reproduce, via the Schr\"{o}d-inger Robertson uncertainty relation, the desired minimal length. The effects of considering GUP are important in systems with energies near the Planck scale. A particularly relevant example of such systems is the very early universe, in which quantum effects of gravity are expected to be dominant. It has also been argued that it is possible to apply the GUP concepts to others more common systems with scales different from the Planck scale, as it was done for example in \cite{AKHOURY200337}. The adimensional GUP parameter $\alpha_0$ in Eq. (\ref{Eqq.GUP.1}) is often taken to be of order one in theoretical calculations, such that the modification to the uncertainty principle only becomes relevant at the Planck scale. Phenomenologically the usual choice is to set $\alpha_0>0$, since in various derivations and thought experiments of GUP it seems a reasonable election. However in some situations the GUP parameter is chosen to be negative ($\alpha_0<0$). In principle, positive and negative GUP parameters are allowed. Each one of them is associated with diverse phenomena that are not in general related among them. In \cite{ONG2018217} the author argues this fact by saying that GUP is largely heuristic, and even as a phenomenological model one can explore various possibilities of its "parameter space" in search of viable options that provides sensible physics. As mentioned the GUP proposal can be understood as originating from Gedanken-experiments of micro black-hole scattering  \cite{PhysRevD.81.084030}, and was also obtained in the framework of string theory. In this work we will able to derive GUP explicitly from the modified statistics due to $S_+$ and $S_-$, having as a consequence a negative and a positive GUP deformation parameter respectively.

Since the seminal work by Shannon \cite{Shannon1948} about information entropy to quantify predictability in a stochastic process, several other measures of information have been proposed in the literature \cite{Renyi1970,PhysRevE.66.056125,Tsallis1988,ABE1997326,Gorban2010}. By maximizing these information measures \cite{AbeBeckCohen2007}, their corresponding probability distributions can be calculated. Some of these generalized information and entropy measures and their potential physical applications have been discussed elsewhere \cite{Beck2009}. In their  work Beck and Cohen \cite{BECK2003267,StraetenBeck2008} considered nonequilibrium systems with a long-term stationary state that possess a spatio-temporally fluctuating intensive quantity. They have shown that after averaging over the fluctuations one can obtain non-extensive statistical mechanics \cite{TSALLIS1998534}. This procedure allows to calculate the corresponding effective Boltzmann factor $B(E)$, from it, the corresponding generalized entropy can be  can be obtained \cite{PhysRevE.67.026106}. The analysis of these $B(E)$ showed that all these statistics present the same behavior for a small variance of the fluctuations \cite{BECK2003267}. An extended discussion exists in the literature analyzing the possible viability of these kind of models to explain several physical phenomena \cite{Beck2009,PhysRevLett.84.2770,doi:10.1143/JPSJ.70.3247}.

Different entropy functionals have been explored to describe non-equilibrium phenomena, most of them are non-extensive entropies. In particular, in  \cite{e12092067} entropies that only depend on the probabilities have been proposed. These entropies are obtained considering a $\Gamma$ (or $\chi^2$) inverse temperature $\beta$ distribution depending on a parameter $p_l$, to be identified with the probability associated with the microscopic configuration of the system, and following the procedure in \cite{PhysRevE.67.026106,e12092067}, the corresponding modified entropies are calculated, named $S_\pm$. One can consider, as mentioned, several distributions, but all possible entropies functions of the probabilities will agree with $S_\pm$ up the first two terms in their expansion, the next terms are very small, and are nearly the same for all possible generalized entropies, so that, $S_\pm$ basically represent the whole family of generalized entropies depending only on the probability \cite{PhysRevE.88.062146}.

The question arises whether modifications to the entropy could also affect the Heisenberg uncertainty relations. There are clues that this is the case in the modified quantum equations obtained from non-extensive statistics \cite{PhysRevD.105.L121501,nobretsallis}. Also the concept of Quantropy which deepens the correspondence between quantum mechanics and statistical mechanics gives light to this connection. The corrections to the BG statistics given by $S_{\pm}$ entropies are relevant in the quantum regime, as can be seen in \cite{obregoncabo}, \cite{PhysRevD.95.124031} and \cite{MARTINEZMERINO2022137085}. In order to follow \cite{coles15}, we would need modified Fourier relations, which imply generalized exponentials and logarithms. However here we would take a different approach, in which we construct an effective Hamiltonian from the non-extensive probability distribution, in the same way, we define an effective momentum in terms of the momentum at low energies or usual momentum. Using the effective momentum we find a modified commutation relation, which implies a modified uncertainty relation, obtained from entropic principles. In the work \cite{obregon-torres-gil} it was shown that statistics associated to $S_+$ and $S_-$ have different characteristics. The first entropy gives rise to an effective potential related with an effective repulsive contribution term, and the second one gives rise to an effective attractive contribution.

There exists an ample exploration of the Generalized Uncertainty Principle in the literature, with origins in quantum gravity theories \cite{GROSS1988407,doi:10.1142/S021827180200302X,CAPOZZIELLO}.  As mentioned it is possible to obtain the Heisenberg uncertainty relation from entropic principles.  Now we question whether it is possible to obtain a generalized uncertainty relation, or what is the same, a modified Heisenberg algebra when considering generalized entropies. 

The structure of this Letter is as follows: In section \ref{sec:Modified_commutation_relation_because_of_GUP}  we discuss briefly the principal characteristics of the Generalized Uncertainty Principle (GUP). In section \ref{sec:Logarithms_and_exponentials_generalized}  we review the basic concepts associated with the modified distributions for $S_+$ and $S_-$ and define the generalized exponentials associated with the non-extensive probability distribution. In section \ref{sec:The_effective_Hamiltonian} we find an effective Hamiltonians and employ them to determine modified commutation relations between the coordinates and the high-energy momenta. In section \ref{Sec:Discussions and conclusions} we summarize our work and comment on potential applications.

\section{Modified commutation relations due to GUP}\label{sec:Modified_commutation_relation_because_of_GUP}
The description of a minimal length has been elaborated in various forms: as a modification of the uncertainty relation without the requirement of a particular representation for the corresponding quantum operators \cite{SCARDIGLI199939,scardigli2015gravitational,CASADIO2020135558}; as a consequence a modification of classical mechanics arises \cite{Mignemi_2012,doi:10.1142/S0217751X13501315,PhysRevD.90.105027,doi:10.1142/S0218271820500704,Bosso.PhysRevD.97.126010} due to the modified position-momentum commutation relations \cite{MAGGIORE199383,PhysRevD.52.1108,PhysRevD.84.044013}. In this work, we will focus on this last approach. In \cite{Bosso2021}, the case of a generic commutation relation between operators $\hat{p}$ and $\hat{q}$ in one dimension is considered:
\begin{eqnarray}
\left[\hat{q},\hat{p}\right]=if\left(\hat{p}\right).\label{eq:Arbitrary_Commut_Relati}
\end{eqnarray}
One first aspect to notice is that, since the commutator of two observables is anti-Hermitian, the function $f$, when regarded as a function of a real variable, has real values.

In this paper, we consider the momentum representation of the position and momentum operators compatible with the commutator in Eq. (\ref{eq:Arbitrary_Commut_Relati})
\begin{eqnarray}
\hat{q}=\hat{x}=i\frac{d}{dk},\quad\quad\quad\quad\quad \hat{p}=p(k).
\end{eqnarray}
In order to implement the notion of a minimal uncertainty or minimal length $l_{Pl}$, let us now suppose that one can increase $p$ arbitrarily, but that $k$ has an upper bound. This effect will show up when $p$ approaches a certain scale $M_{Pl}$, that it is naturally the Planck scale. The physical interpretation of this is that particles can not possess arbitrarily small Compton wavelengths $\lambda = 2\pi/k$ and that arbitrarily small scales could not be resolved anymore \cite{HOSSENFELDER200385}.

To incorporate this behavior, we assume a relation $k = k(p)$, which can also be written as $p=p(k)$, between $p$ and $k$. The quantization of these relations is straightforward. The commutators between $\hat{k}$ and $\hat{x}$ remain in the standard form given by $\left[\hat{x},\hat{k}\right]=i$. Inserting the functional relation between the wave vector and the momentum then yields the modified commutator. In the momentum representation, we have \cite{PhysRevD.94.123505}
\begin{eqnarray}
\left[\hat{x},\hat{p}\right]=\left[i\frac{d}{dk},p(k)\right]=i\frac{d}{dk}p(k).\label{Eqq.General.Commut.Relat}
\end{eqnarray}
This results in the generalized uncertainty relation
\begin{eqnarray}
\Delta x\Delta p\geq\frac{1}{2}\left|\left\langle\frac{dp(k)}{dk}\right\rangle\right|.
\end{eqnarray}
Comparing Eq. (\ref{eq:Arbitrary_Commut_Relati}) with Eq. (\ref{Eqq.General.Commut.Relat}) we identify
\begin{eqnarray}
f(p)=\frac{d}{dk}p(k),
\end{eqnarray}
is easy to see that $k$ is related to the physical momentum through
\begin{eqnarray}
k(p)=\int_0^\infty\frac{dp}{f(p)}.
\end{eqnarray}
The same relationship has been obtained in \cite{HOSSENFELDER200385}.

Comparing Eq. (\ref{Eqq.General.Commut.Relat}) and Eq. (\ref{Eqq.GCR.1}) we find that Eq. (\ref{Eqq.General.Commut.Relat}) translates to a differential equation
\begin{eqnarray}
\frac{d}{dk}p(k)=\left(1+\alpha\ p^2\right),
\end{eqnarray}
and solving this, one obtains a functional relation between high-energy and low-energy momenta, $p$ and $k$ respectively \cite{PhysRevD.94.123505}
\begin{eqnarray}
p(k)=\frac{\tan \left(\sqrt{\alpha }\  k\right)}{\sqrt{\alpha }},\label{Eqq.p.in.function.k.Tan}
\end{eqnarray}
expanding $\tan(\sqrt{\alpha }\  k)$ and approximating to the first order in $\alpha$, we assume that the terms $\mathcal{O}(\alpha^2)$ are much smaller in magnitude in comparison to terms $\mathcal{O}(\alpha)$ \cite{MAJUMDER2012291}. These enables us to express Eq. (\ref{Eqq.p.in.function.k.Tan}) as a series in terms of low-energy momentum $k$ \cite{BISHOP2020135209,https://doi.org/10.48550/arxiv.2206.05064}
\begin{eqnarray}
p=k+\frac{\alpha}{3} k^3\cdots,\label{Eqq.Eff.Momen.From.GUP}
\end{eqnarray}
so, Eq. (\ref{Eqq.GCR.1}) can be write as follows 
\begin{eqnarray}
\left[\hat{x},\hat{p}(k)\right]=i\left(1+\alpha\  k^2+\cdots\right).\label{Eqq.Modd.Conm. Relat.Usual.GUP}
\end{eqnarray}
Precision data available on the energy levels can be used to constrain the scale of the new physics, which should be the quantum gravity scale. Writing, $\alpha=\alpha_0/M_{Pl}^2$, $\alpha_0$ is a dimensionless constant. The Planck mass scale is denoted by $M_{Pl}$, and the parameter $\alpha_0$ is of the order of unity. This parameter can be positive or negative, both possibilities have equal dignity. The structure of the models with positive and negative parameters usually differs. For example, while in the first case, one usually has a minimal DeBroglie wavelength and no constraint on the physical momentum, in the second case on has a maximum physical momentum and no constraint on the DeBroglie wave length.

A different modification in the Heisenberg algebra is presented in \cite{Bosso_2021}. In order to find the effective GUP-modified dynamics for a black hole interior, a modified algebra inspired by GUP is  proposed. This affects the classical algebra of the dynamical variables, and it removes the singularity of the Schwarzschild black hole, for a negative deformation parameter. The Doubly Special Relativity (DSR) theory on the other hand, also suggest a modification to the commutators. These commutators, which are consistent with string theory, black holes physics, and DSR, are described in \cite{PhysRevD.84.044013}. In \cite{doi:10.1142/S0219887822500979}, other different modifications of the Heisenberg Uncertainty Principle are presented and implemented in a Bianchi I model.

\section{Superstatistics and modified Entropies depending only on the probability}
\label{sec:Logarithms_and_exponentials_generalized}
In this section, we describe the origin of the entropies $S_{\pm}$, their connection with Superstatistics and their functional form. Furthermore, we will define generalized functions related to them, as logarithms and exponentials.

Let us start by considering systems that can be described in the frame of Superstatistics \cite{beckcohen}. These are systems that are composed by constituents in local equilibrium at a temperature $T=\frac{1}{k\beta}$ with a given
distribution $\Gamma(\beta)$. Those distributions give rise to
different modified entropies such as Tsallis, R\'enyi, etc.,
and in particular they also give rise to parameter independent entropies that we will denote $S_{\pm}$.

Here we will consider the last ones, which posses a $\Gamma$ (or $\chi^2$) inverse temperature $\beta$ distribution depending on a parameter $p_l$, to be identified with the probability associated with the microscopic configuration of the system. We can write these parameter $p_l-\Gamma$ distribution as
\begin{eqnarray}
f_{p_1}\left(\beta\right)=\frac{1}{\beta_0p_l\Gamma\left(\frac{1}{p_l}\right)}\left(\frac{\beta}{p_l\beta_0}\right)^{\frac{1-p_l}{p_l}}e^{-\beta/p_l\beta_0},\label{Gamm.Distr.p_l}
\end{eqnarray}
where $\beta_0$ is the average inverse temperature. From it, one can get an effective Boltzmann factor form
\begin{eqnarray}
B(E)=\int_0^{\infty}d\beta\ f(\beta)e^{-\beta E},\label{Eqq.Eff.Boltz.Factor}
\end{eqnarray}
where $E$ is the energy of a microstate associated with each of the considered cells. The ordinary Boltzmann factor is recovered for $f(\beta) = \delta(\beta-\beta_0)$.  By performing an integration over $\beta$ in Eq. (\ref{Eqq.Eff.Boltz.Factor}), one yields the generalized Boltzmann factor
\begin{eqnarray}
B_{p_l}(E)=\left(1+p_l \beta_0 E\right)^{-\frac{1}{p_l}}.\label{Gen.Boltz.Fact.p_l}
\end{eqnarray}
This expression can be expanded for small $p_l\beta_0 E$, to get
\begin{eqnarray}
B_{p_l}(E)=e^{-\beta_0 E}\left[1+\frac{1}{2}p_l\beta_0^2E^2-\frac{1}{3}p_l^2\beta_0^3E^3+\cdots\right].
\end{eqnarray}
Following \cite{PhysRevE.67.026106,e12092067}, we present the procedure to obtain the entropy corresponding to the $f (\beta)$ distribution Eq. (\ref{Gamm.Distr.p_l}) and to its associated generalized Boltzmann factor Eq. (\ref{Gen.Boltz.Fact.p_l}). We begin by defining the entropy $S=k\sum_{l=1}^{\Omega}s(p_l)$ in terms of a generic function $s(p_l)$, where $p_l$ can be considered at this moment an arbitrary parameter; considering for $s(x)=-x\ln x$ the Shannon entropy is recovered. As shown in \cite{PhysRevE.67.026106} it is possible to express $s(x)$ and a generic internal energy $u(x)$ in terms of integrals of a function $E(y)$ that is obtained from the Boltzmann factor of interest $B(E)$ . By these means $s(x)$ and $u(x)$ can be written as
\begin{eqnarray}
s(x)=\int_0^x\frac{\alpha+E(y)}{1-\frac{E(y)}{E^*}}dy
\end{eqnarray}
and
\begin{eqnarray}
u(x)=\left(1+\frac{\alpha}{E^*}\right)\int_0^x\frac{dy}{1-\frac{E(y)}{E^*}},
\end{eqnarray}
where $E(y)$ is to be identified with the inverse function of $B_{p_l}(E)/\int_0^\infty dE'B_{p_l}(E')$. In our case, the starting points are the distribution Eq. (\ref{Gamm.Distr.p_l}) and the Boltzmann factor Eq. (\ref{Gen.Boltz.Fact.p_l}). $E(y)$ and $E^*$ are given by
\begin{eqnarray}
E(y)=\frac{y^{-x}-1}{x},\quad\quad\quad\quad\quad E^*=-\frac{1}{x}.
\end{eqnarray}
A straightforward calculation gives for $u(x)$ and $s(x)$
\begin{eqnarray}
u(x)=x^{x+1},\quad\quad\quad\quad\quad\quad\quad s(x)=1-x^x.
\end{eqnarray}
By these means the entropy results in
\begin{eqnarray}
S_{+}=k\sum_{l}\left(1-p_l^{p_l}\right),\label{Eqq.Entropy.S_+}
\end{eqnarray}
where $k$ is the conventional constant and $\sum_{l=1}^\Omega p_l=1$. If we change appropriately $p_l$ by $-p_l$  in Eq. (\ref{Gamm.Distr.p_l}), another entropy can be obtained, and following the same procedure that was used to find $S_+$ we obtain
\begin{eqnarray}
S_{-}=k\sum_{l}^{\Omega}\left(p_l^{-p_l}-1\right).\label{Eqq.Entropy.S_-}
\end{eqnarray}
The expansion of Eq.(\ref{Eqq.Entropy.S_+}) gives
\begin{equation}
    -\frac{S_+}{k}=\sum_{l=1}^{\Omega}\left[p_l\ln{p_l}+\frac{\left(p_l\ln{p_l}\right)^2}{2!}+\frac{\left(p_l\ln{p_l}\right)^3}{3!}+\cdots\right],
\end{equation}
where the first term in the expansion constitutes the Shannon entropy. For the case of euqal probabilities $p_l=1/\Omega$, in terms of the Boltzmann-Gibbs (Shannon) entropy $S_B=k\ln{\Omega}$, we can write
\begin{align}
     \frac{S_{+}}{k}=&\frac{S_B}{k}-\frac{1}{2!} \left(\frac{S_B}{k}\right)^2 \exp \left(-\frac{S_B}{k}\right)\\
    +&\frac{1}{3!} \left(\frac{S_B}{k}\right)^3 \exp \left(-2\frac{S_B}{k}\right)+\cdots.\label{S+.Entro.SB}
\end{align}
Also one can perform the analogous expansion for $S_-$
\begin{align}
     \frac{S_{-}}{k_\beta}=&\frac{S_B}{k_\beta}+\frac{1}{2!} \left(\frac{S_B}{k_\beta}\right)^2 \exp \left(-\frac{S_B}{k_\beta}\right)\\
    +&\frac{1}{3!} \left(\frac{S_B}{k_\beta}\right)^3 \exp \left(-2\frac{S_B}{k_\beta}\right)+\cdots.\label{S-.Entro.SB}
\end{align}
From Eqs. (\ref{S+.Entro.SB}, \ref{S-.Entro.SB}) we can see those corrections due to the non-extensivity are different for each entropy (the sign of correction changes).\par
Using expression (\ref{Eqq.Entropy.S_+}), the corresponding functional including restrictions is given by
\begin{equation}
    \Phi_+=\frac{S_+}{k}-\gamma\sum_{l=1}^{\Omega}p_l-\beta\sum_{l=1}^{\Omega}p_l^{p_l+1}E_l,
\end{equation}
where the second restriction concerns the average value of the energy and $\gamma$ and $\beta$ are Lagrange parameters, and then by maximizing $\Phi_+$, $p_l$ is obtained for $S_+$ as
\begin{equation}
    1+\ln{p_l}+\beta E_l\left(1+p_l+p_l\ln{p_l}\right)=p_l^{-p_l}.
\end{equation}
And in a similar way, constructing and maximizing the functional $\Phi_-$ for $S_{-}$ one has
\begin{equation}
    1+\ln{p_l}+\beta E_l\left(1-p_l-p_l\ln{p_l}\right)=p_l^{p_l}.
\end{equation}
The dominant term in these expressions corresponds to the Gibbs-Boltzmann prediction, $p_l=e^{-\beta_0E_l}$. In general, however, we cannot analytically express $p_l$ as a function of $\beta E_l$.

We write then the two nonparametric entropies in terms of generalized logarithms $\log^\pm$ \cite{fuentes2021generalised}
\begin{eqnarray}
S_{\pm}(X)&=&-\sum_{i}^{N}p(x_i)\log^{\pm}(p(x_i)).\label{Eqq.Entropy.S_+.S_-}
\end{eqnarray}
where the states $x_1,...,x_N$ give rise to the probabilities $p(x_1),...,p(x_N)$.
\begin{eqnarray}
\log^+(x)&=&-\frac{1-x^x}{x}\\
\log^-(x)&=&-\frac{x^{-x}-1}{x},
\end{eqnarray}
From such definitions, it becomes clear that the functions $ \log^\pm $ do not comply with the three laws of logarithms. Additionally, the corresponding inverse functions of the generalized logarithms do not have a closed-form, however they do posses solutions in series. We can also consider a numerical representation of these functions. These functions have been built as \footnote{This Ansatz is a variation over the usual probability distribution. Also as it was mentioned earlier, with this Ansatz one can find a recurrent series solution of the generalized entropy constraint equations, with a given radius of convergence. Therefore we also propose the Ansatz as a fit to the constraint equations, which is valid in certain region, and offers a small mean square deviation from the exact probability.}
\begin{eqnarray}
\exp_\pm(-x)&=&\exp(-x)\sum_{j=0}^{\infty}a^\pm_jx^j,\label{GeneralizedExponential}
\end{eqnarray}
and the value of the first coefficients $ a^\pm_j $ are presented in the Table \ref{tab:1}.

\begin{table}[!htb]
    \centering
    \begin{tabular}{ccc}
 &$a^+_j$&$a^-_j$\\\hline
j = 4 & 1.284852  & 0.893692 \\
j = 3 & -1.205053    & 0.851734 \\
j = 2 & 0.747398     & -0.586262   \\
j = 1 & 0.000029   & -0.333335  \\
j = 0 & 1            & 1          \\\hline
\end{tabular}
    \caption{List of first values of the coefficients $ a^\pm_j $ from the expansion (\ref{GeneralizedExponential}). This is obtained as a fit till reaching convergence and as a recurrent series expansion.}
    \label{tab:1}
\end{table}
Associated with the exponential Eq. (\ref{GeneralizedExponential}) one can write the corresponding probabilities.

\section{The effective Hamiltonian}
\label{sec:The_effective_Hamiltonian}
In the following, we find the effective Hamiltonians $H_{\pm}$ from the non-extensive probabilities and from them we obtain the high-energy momenta $p_{\pm}$.  With these effective momenta we can compute the modified commutation relation between them and position in the case of a one-dimensional system. The energetic scale in the Hamiltonian is $M_{Pl}$, such that $H=M_{Pl} \bar{H}$ and the dimensions of $H$ and $\bar H$ are $[H]=M_{Pl}$
and $[\bar H]=1$. In many works, the notation $\bar{H}$ represents the average value of an observable $H$, but in this work denotes a dimensionless quantity. We are considering the light velocity as $c=1$, therefore, $M_{Pl}$ is a mass parameter.

The probability of a state is given in terms of the dimensionless Hamiltonian by
\begin{eqnarray}
P=e^{-\bar{H}}=e^{-H/M_{Pl}},
\end{eqnarray}

and from Eq. (\ref{GeneralizedExponential})
\begin{eqnarray}
P_\pm=\exp(-\bar{H})\sum_{j=0}^{\infty}\bar{a}^\pm_j\bar{H}^j,\label{6}
\end{eqnarray}
the coefficients $[{a}^\pm_j]=1$ are also dimensionless. On the other hand, we consider as an Ansatz that the effects of different statistics in a given system can now be cast by means of the  effective Hamiltonian \cite{obregon-torres-gil}
\begin{eqnarray}
P_\pm=\exp\left(-\bar H_\pm\right)=\exp(-\bar{H})\sum_{j=0}^{\infty}a^\pm_j\bar{H}^j,\label{Anzats.Probabil}
\end{eqnarray}
being $\bar H_\pm$ the dimensionless  effective Hamiltonian, $[\bar{H}_{\pm}]=1$. Using the Ansatz in Eq. (\ref{Anzats.Probabil}) the effective Hamiltonian can be expressed in the terms of the low energy dimensionless Hamiltonian $\bar{H}=\frac{1}{2}\bar{k}^2+\bar{V}(x)$, where $\bar{k}$ and $\bar{V}$ are dimensionless quantities. Now, if we consider for simplicity the potential function $V(\hat{x})=0$, one obtains
\begin{multline}
\bar{H}_{\pm}=\frac{1}{2}\left(1-{a}^\pm_1\right)\bar{k}^2+\frac{1}{8}\left(\left(a^\pm_1\right)^2-2{a}^\pm_2\right)\bar{k}^4+\\
\frac{1}{24}\left(-\left(a^\pm_1\right)^3+3 {a}^\pm_1 {a}^\pm_2-3{a}^\pm_3\right)\bar{k}^6+\cdots
\end{multline}
On the other hand, we consider that the effective Hamiltonian $\bar{H}_{\pm}$ can be written in terms of an effective momentum $\bar{p}_\pm$, in the usual way
\begin{eqnarray}
\bar{H}_{\pm}=\frac{\bar{p}_\pm^2}{2},\label{Eff.Hamil.In.Funct.Eff.Momen}
\end{eqnarray}
these momenta are functions of $\bar{k}$.  Remembering our notation $\bar{x}=M_{Pl}x$ and $\bar{p}_\pm=p_\pm/M_{Pl}$, it is then possible to express the generalization for the momentum in its full dimensional form
\begin{eqnarray}
p_\pm=k+\frac{\alpha_\pm}{3} k^3+\cdots,\label{Eqq.Eff.Moment.S_{+-}}
\end{eqnarray}
where, we denote $\alpha_\pm=\alpha_\pm^{0}/M_{Pl}^2$, and
\begin{eqnarray}
\alpha_\pm^{0}=\frac{3 \left(\left(a^\pm_1\right)^2-2{a}^\pm_2\right)}{8 \left(1-{a}^\pm_1\right)}.\label{Eq.Dimenssles.Deford.Paramter}
\end{eqnarray}
In Eq. (\ref{Eqq.Eff.Moment.S_{+-}}), we adopt a more convenient normalization for the momentum, such that the coefficient in the first term is 1. In models of GUP, the existence of a minimal length scale leads to the generalization of the momentum operator \cite{nozari2006some,HOSSENFELDER200385,Brau_1999}, as in Eq. (\ref{Eqq.Eff.Momen.From.GUP}). In contrast, a similar generalization is obtained in Eq. (\ref{Eqq.Eff.Moment.S_{+-}}), starting from non-extensive entropies.\par

The commutation relation between $\hat{p}_\pm$ and $\hat{x}$ can be calculated using Eq. (\ref{Eqq.Eff.Moment.S_{+-}}), and the usual commutation relation $\left[\hat{x},\hat{k}\right]=i$ ($\hbar=1$)
\begin{eqnarray}
\left[\hat{ x},\hat{p}_\pm\right]=i \left(1+\alpha_\pm {k}^2+\cdots\right),\label{Modif.Conmut.Relat.Diment.x.p_pm}
\end{eqnarray}
taking into account that $\alpha_\pm p_{\pm}^2=\alpha_\pm k^2+\cdots$, it allows us to express Eq. (\ref{Modif.Conmut.Relat.Diment.x.p_pm}) as
\begin{eqnarray}
\left[\hat{ x},\hat{p}_\pm\right]=i \left(1+\alpha_\pm\ \hat p_\pm^2\right),\label{Modif.Conmut.Relat.Diment.x.p_pm.Fam}
\end{eqnarray}
this modified commutation relation is the counterpart of Eq. (\ref{Eqq.GCR.1}). We conclude that the modified entropies deform the Heisenberg commutation relation, i.e. they have as a straightforward consequence the GUP. Thus, Eq. (\ref{Modif.Conmut.Relat.Diment.x.p_pm.Fam}) represents a modified Heisenberg algebra similar to Eq. (\ref{Eqq.GCR.1}), with the difference that Eq. (\ref{Modif.Conmut.Relat.Diment.x.p_pm.Fam}) is obtained from entropic principles and Eq. (\ref{Eqq.GCR.1}) is a phenomenological model proposed in quantum gravity theories.

From Eq. (\ref{Eq.Dimenssles.Deford.Paramter}) and using the values in Table \ref{tab:1} we note that for the considered statistics one has positive (negative) values of $\alpha_-(\alpha_+)$ for $S_-(S_+)$,  i.e,  $\alpha^0_+=-0.560565$ and $\alpha^0_{-}=0.361022$. In GUP, the deformation parameter can have positive and negative values, in each case, for a considerable group of physical phenomena. \cite{PhysRevD.98.126018,scardigli2015gravitational,Ong_2018,ONG2018217,ong2018effective,universe8070349,Bosso_2021}. In our work both cases are derived, this is because the entropies $S_+$ and $S_-$ are different in their origin. Therefore, different probability distributions associated with them arise. Hence the quantum corrections will be different.  It is of interest to notice that for example while in the first case, one usually has a minimum deBroglie wavelength and no constraint on the physical momentum, in the second one gets a maximum physical momentum and no constraint on the de Broglie wave length. Given that the $\alpha$ parameters obtained are associated with quantum gravity effects, one would expect to have a minimal wave length ($\alpha_-$) or in the other case a maximum momentum ($\alpha_+$) closely related to the Planck scale ($l_{Pl}$ or $1/l_{Pl}$). These values of the parameters denoting the correction terms to the uncertainty principle are relevant for extremely high energies, namely in the realm of the quantum gravity effects.

 Following a similar procedure, we also found GUP expressions for the R\'enyi and Tsallis entropies. In particular, for the Tsallis statistics one has $\alpha_q=(1-q)/M_{Pl}^2$ \footnote{Notice that the procedure employed can be applied to any statistics where coefficients $a_1$ and $a_2$ are determined for the probability distribution.}. Notice that the $\alpha_q$ can be also negative or positive.

As stated, the Heisenberg uncertainty principle is a direct consequence of the entropic uncertainty principle.
As shown here GUP emerges in a completely similar manner from generalized entropies $S_+$, $S_-$ and $S_q$. So GUP is a direct consequence of non-extensive entropies, and one would then expect that probabilities on the GUP formalism will be related with the modified entropy probabilities.

\section{Discussions and conclusions}\label{Sec:Discussions and conclusions}
As already seen, from the entropic uncertainty relation, the Heisenberg uncertainty relation is obtained, when considering a Gaussian probability distribution. On the other hand, it has been obtained that the Heisenberg uncertainty relation is modified to consider quantum gravity effects Eq. (\ref{Eqq.GUP.1}). This modification is commonly known as GUP.  This fact motivated us to think that GUP could have as a possible origin  modified entropies. As GUP is related with string scattering \cite{PhysRevD.81.084030} and/or a kind of diffraction of micro black holes \cite{SCARDIGLI199939}, the natural scale is the Planck one, i.e. the quantum gravity scale. Then a modified statistics seems to be a signature of quantum gravity phenomena. We decided to explore generalized entropies that could lead us to a GUP.

We paid attention to the modified entropies that depend only on the probability distributions Eq. (\ref{Eqq.Entropy.S_+.S_-}). From these generalized entropies, we obtain an effective momentum $p_\pm$ depending on the usual momentum $k$, as can be seen from the Eq. (\ref{Eqq.Eff.Moment.S_{+-}}). This effective momentum is similar to the one obtained in equation Eq. (\ref{Eqq.Eff.Momen.From.GUP}) in the usual GUP. Therefore, when calculating the commutator between $p_\pm$ and the low energy position $x$  a modified Heisenberg commutation relation arises Eq. (\ref{Modif.Conmut.Relat.Diment.x.p_pm.Fam}). These modified relations are of the same kind as the ones obtained in the standard GUP formalism Eq. (\ref{Eqq.GCR.1}). In this way, we show that it is possible to derive GUP from an entropic principle. As stated in Table \ref{tab:1} we have found the deformation parameters associated with both statistics ($\alpha^0_+<0$ and $\alpha^0_{-}>0$). Actually the quantum corrections due to both probability distributions associated with $S_\pm$ are different. This difference could be related to previously observed physical implications of $S_+$ and $S_-$. For example $S_+$ statistics is equivalent to standard statistics with an effective repulsive potential and $S_-$ gives rise to an effective attractive potential \cite{obregon-torres-gil}. This example shows also us that positive and negative GUP parameters have as a consequence a different physical behavior for the same starting physical system. Also when the value of the GUP parameter is positive, we have minimal uncertainty in the position and no constraint on the physical momentum, on the other hand, when it is negative, one gets a maximum physical momentum but no minimal uncertainty in the position. Let us emphasize that the GUP obtained from the generalized entropies describes effects that are relevant at the Planck scale, as seen in the values of the parameters $\alpha_\pm=\alpha_\pm^{0}/M_{Pl}^2$; so they are in the realm of quantum gravity.

\section{Acknowledgements}
OO thanks the support of CONACYT Project 257919 "Teor\'{\i}a M-Topol\'ogica y Teor\'{\i}a de Matrices: Su relaci\'on con invariantes topol\'ogicos, gravitaci\'on y cosmolog\'{\i}a", University of Guanajuato Projects CIIC 188/2019 and the suppport by PRODEP. NGCB would like to thank the support of CONACyT Project A-1-S-37752, "Teor\'ias efectivas de cuerdas y sus aplicaciones a la f\'isica de part\'iculas y cosmolog\'ia", UG Project CIIC 148/2021 "Exploraci\'on del paisaje de la teor\'ia de cuerdas: geometr\'ia, dualidades y aprendizaje de m\'aquina", UG Project CIIC 264/2022 "Geometr\'{\i}a de dimensiones extras en teor\'{\i}a de cuerdas y sus aplicaciones f\'{\i}sicas" and PRODEP. WY would like to thank the support granted by CONACYT.


\bibliographystyle{JHEP}
\bibliography{References}

\end{document}